\documentclass[aps,prd,preprint,showkeys,superscriptaddress]{revtex4}

\usepackage{latexsym}
\usepackage{amsmath,amssymb}
\usepackage{graphicx}
\usepackage{subfigure}
\usepackage{xcolor}

\begin{document}


\title{Local thermodynamics of KS black hole}

\author{Myungseok Eune}
\email[]{younms@sogang.ac.kr}
\affiliation{Research Institute for Basic Science, Sogang University, Seoul, 121-742, Korea}

\author{Bogeun Gwak}
\email[]{rasenis@sogang.ac.kr}
\affiliation{Research Institute for Basic Science, Sogang University, Seoul, 121-742, Korea}
\affiliation{Center for Quantum Spacetime, Sogang University, Seoul 121-742, Korea}

\author{Wontae Kim}
\email[]{wtkim@sogang.ac.kr}
\affiliation{Research Institute for Basic Science, Sogang University, Seoul, 121-742, Korea}
\affiliation{Center for Quantum Spacetime, Sogang University, Seoul 121-742, Korea}
\affiliation{Department of Physics, Sogang University, Seoul 121-742, Korea}

\date{\today}

\begin{abstract}
  We study the thermodynamics of the KS black hole which is an
  asymptotically flat solution of the HL gravity. In particular, we
  introduce a cavity to describe the thermodynamics at a finite
  isothermal surface on general ground and to get a well-defined
  thermodynamics. We show that there exists a locally stable small
  black hole which tunnels to the hot flat space below the critical
  temperature and to the large black hole above the critical
  temperature. Moreover, it turns out that the remnant decays into the
  vacuum through a quantum tunneling.
\end{abstract}


\maketitle

\section{Introduction}
\label{sec:intro}

It has been claimed that the intrinsic entropy of a black hole is
proportional to the area of the event horizon by
Bekenstein~\cite{Bekenstein:1972tm}. Subsequently, Hawking has shown
that the black hole has thermal radiation through a quantum field
theoretical analysis~\cite{Hawking:1974sw} and studied the
thermodynamic phase transition~\cite{Hawking:1982dh}. In particular,
the thermodynamics of black holes has been also studied in a cavity to
get well-defined canonical ensembles~\cite{Allen:1984bp, York:1986it,
  Whiting:1988qr, Brown:1994gs, Creighton:1995uj, Dehghani:2001af,
  Zaslavskii:2003cr, Grumiller:2007ju, Astefanesei:2009wi}. In the
quasilocal thermodynamics of a black hole, the thermodynamical
quantities such as the energy, the temperature, and so on, are related
to the size of cavity, whereas the entropy is not since the entropy
can be regarded as a conserved Noether charge~\cite{Wald:1993nt}.

On the other hand, Ho\v{r}ava has recently proposed the
Ho\v{r}ava-Lifshitz (HL) gravity~\cite{Horava:2008ih, Horava:2009uw},
inspired by condensed matter models of dynamical critical
systems~\cite{lifshitz}. The HL gravity is a power-counting
renormalizable theory of gravity with anisotropic scaling of space and
time. The scale transformations of time and space are given by $t \to
b^z t$ and $x^i \to b x^i$, respectively, where $i = 1, \cdots, D$ is
the spatial index, $D$ is the dimension of space, and the Lifshitz
index $z$ is the ``critical exponent'' in the Lifshitz scalar field
theory. One can expect that it recovers the general relativity in the
IR region whereas it becomes a nonrelativistic gravity in the UV
region. The static and spherically symmetric black hole solutions in
the HL gravity have been investigated~\cite{Lu:2009em,
  Kehagias:2009is}. While some of solutions give the asymptotically
anti-de Sitter spacetime~\cite{Lu:2009em}, the black hole solution
called the KS solution \cite{Kehagias:2009is} behaves like the
Schwarzschild metric at infinity. Then, there have been many studies
for the entropy of these black holes~\cite{Cai:2009qs, Myung:2010dv,
  Eune:2010kx, Radinschi:2010iw, Myung:2009us, Liu:2011zzm,
  Biswas:2011gr, Khatua:2011sh, Liu:2011zzu}. Especially, the ADM mass
for the KS black hole was identified with a parameter from the
asymptotic expansion of the metric.  Moreover, it has been shown that
the entropy is modified by the logarithmic term from the first law of
thermodynamics~\cite{Myung:2009us}. This logarithmic correction to the
entropy can be also obtained from the quantum tunneling
method~\cite{Parikh:1999mf, Liu:2011zzu}.

In this paper, we would like to study the thermodynamics of the KS
black hole using the logarithmic corrected entropy in the cavity.  It
should give the expected Hawking-Page type phase transition in the
large black hole like the ordinary thermodynamics of the Schwarzschild
black hole since the KS metric is asymptotically same with the
Schwarzschild metric~\cite{Hawking:1982dh}.  However, for the small
black hole, its behavior will be different from that of the
Schwarzschild black hole.  We find that there exists a locally stable
small black hole which decays into the thermal state without the black
hole below a critical temperature while it can decay into the large
stable black hole above a critical temperature.  Moreover, it will be
shown that the remnant of the lowest mass of the black hole tunnels to
the vacuum.

The paper is organized as follows. In section~\ref{sec:HL}, we
recapitulate the HL gravity and introduce the KS solution.  In
section~\ref{sec:KSterm}, we get the thermodynamic quantities of
Tolman temperature, heat capacity including free energies. In
section~\ref{sec:phenomena}, the thermodynamics and phase transitions
are studied for a given range of the coupling $\omega$. Finally, the
summary and discussion are given in section~\ref{sec:discus}.

\section{Schwarzschild-like black hole in the HL gravity}
\label{sec:HL}

In this section, we would like to recapitulate the HL gravity for our
purposes and introduce a static and spherically symmetric black hole.
Let us start with the four-dimensional line element parametrized as
$ds^2 = -N^2 dt^2 + g_{ij} (dx^i - N^i dt)(dx^j - N^j dt)$, where
$N(t,x^i)$, $N^i(t,x^j)$, and $g_{ij}(t,x^k)$ are the lapse function,
the shift functions, and the spatial metric, respectively.  Along with
the ADM decomposition, the Einstein-Hilbert action is written as
\begin{align}
  S_{\rm EH} = \frac{1}{16\pi G} \int d^4 x \sqrt{g} N \left(K_{ij}
    K^{ij} - K^2 + R - 2\Lambda \right), \label{action:EH}
\end{align}
where $G$ and $\Lambda$ are Newton's constant and cosmological
constant, respectively. The extrinsic curvature $K_{ij}$ is defined by
$K_{ij} = \frac{1}{2N} (\dot{g}_{ij} - \nabla_i N_j - \nabla_j N_i)$.
Here, the dot denotes a derivative with respective to $t$ and
$\nabla_i$ is a covariant derivative with the spatial metric
$g_{ij}$. The action in the HL gravity is given
by~\cite{Horava:2008ih,Horava:2009uw}
\begin{align}
  S_{\rm HL} &= \int dt d^2 {\bf x} \sqrt{g} N
  \bigg[\frac{2}{\kappa^2} (K_{ij} K^{ij} - \lambda K^2) +
  \frac{\kappa^2 \mu^2}{8(1-3\lambda)} \left( \frac{1-4\lambda}{4} R^2
    + \Lambda_W R - 3\Lambda_W^2 \right)
  \notag \\
  & \quad - \frac{\kappa^2}{2 w^4} \left(C_{ij} - \frac{\mu w^2}{2}
    R_{ij} \right) \left(C^{ij} - \frac{\mu w^2}{2} R^{ij} \right)
  \bigg], \label{action:HL}
\end{align}
where $\kappa$, $\mu$, $\lambda$, $w$, and $\Lambda_W$ are constant
parameters, and the Cotton tensor $C_{ij}$ is defined by $C^{ij} =
(\epsilon^{ikl} / \sqrt{g}) \nabla_k \left(R^j_{\phantom{j}l} -
  \frac14 \delta^j_l R \right)$.  In the IR limit, comparing the
action~\eqref{action:HL} to the Einstein-Hilbert
action~\eqref{action:EH}, the speed of light $c$, Newton's constant
$G$, and the cosmological constant $\Lambda$ are identified with $ c =
\frac14 \kappa^2\mu\sqrt{\Lambda_W / (1-3\lambda)}$, $G =
\kappa^2/(32\pi c)$, and $\Lambda = \frac32 \Lambda_W$.

Although the kinetic term in Eq.~\eqref{action:HL} agrees with that of
the Einstein-Hilbert action~\eqref{action:EH} when $\lambda = 1$, the
action~\eqref{action:HL} without $ \Lambda_W$ does not fully recover
Eq.~\eqref{action:EH} in the IR limit. In order to obtain it in the IR
limit, one can deform the HL gravity by introducing a soft violation
term as $\mathcal{L}_V \to \mathcal{L}_V + \sqrt{g} N \mu^2 R$ and
taking the limit of $\Lambda_W$ as $\Lambda_W \to 0$, which is called
the ``deformed HL gravity''. In the deformed HL gravity, the UV
properties are unchanged, whereas there exists a Minkowski vacuum in
the IR limit. In this case, the constants are given by $c =
\kappa\mu^2/\sqrt{2}$, $G = \kappa^2/(32\pi c)$, and $\lambda = 1$.
In the deformed HL gravity, a static and spherically symmetric black
hole is described by the line element~\cite{Kehagias:2009is}
\begin{align}
  ds^2 = -f(r) dt^2 + \frac{dr^2}{f(r)} + r^2 (d\theta^2 +
  \sin^2\theta d\phi^2), \label{metric:KS}
\end{align}
with the metric function $f(r)$ given by
\begin{align}
  f(r) = 1 + \omega r^2 \left(1 - \sqrt{1 + \frac{4M}{\omega r^3}}
  \right), \label{metric:f}
\end{align}
where $\omega = 16\mu^2/\kappa^2$ and $M$ is a positive constant.  The
black hole described by the metric function~\eqref{metric:f} is called
the KS black hole. In the limit of $\omega \to \infty$ or $r \to
\infty$, it can be reduced to the form of $f(r) = 1 - 2M/r +
2M^2/(\omega r^4) + \cdots$, which gives the Schwarzschild metric
function and has the Minkowski vacuum at infinity.  The third term in
the expansion shows that the IR properties of the KS black hole are
different from the Reissner-Nordstr\"om (RN) black hole whose metric
function is given by $f(r) = 1 - 2M/r + Q^2/r^2$, where $Q$ is the
electric charge of the RN black hole. From the metric
function~\eqref{metric:f}, the outer $r_+$ and the inner $r_-$
horizons are given by
\begin{align}
  r_\pm = M \left( 1 \pm \sqrt{1 - \frac{1}{2\omega M^2}}
  \right), \label{horizon}
\end{align}
where the mass is restricted to $M \ge 1/\sqrt{2\omega}$. There exists
the extremal black hole with the horizon $r_e$ and the mass $M_e$
given by $r_e = M_e = 1/\sqrt{2\omega}$.

\section{Thermodynamic Quantities}
\label{sec:KSterm}

We now introduce a cavity with a finite size $r$ in order to obtain
the well-defined thermodynamics for the asymptotically flat black hole
\cite{Whiting:1988qr}. In the line element~\eqref{metric:KS}, there is
the maximum of a mass for a given coupling and a given size $r$,
$M_{\rm max}= (1+2\omega r^2)/(4 \omega r)$ because of $r \ge r_+$.
By the way, from the reality condition of Eq.~\eqref{horizon}, there
exists the minimum of mass $M_{\rm min} = 1/\sqrt{2\omega}$, which is
just the mass of the extremal black hole.  Thus, the black hole mass
should be bounded.

Next, the Tolman temperature $T$ on the boundary of the cavity is
given by~\cite{Tolman:1930zza}
\begin{align}
  \label{eq:tem}
  T=\frac{T_H}{\sqrt{f(r)}},
\end{align}
where the Hawking temperature is $T_H = (-1+2\omega r_+^2)/[8\pi r_+(1
+ \omega r_+^2)]$. In particular, it is interesting to note that the
entropy $S$ of the KS black hole is written as the Bekenstein-Hawking
entropy with the logarithmic term~\cite{Myung:2009us, Liu:2011zzu},
\begin{align}
  \label{bhent}
  S &= \frac{A_H}{4} + \frac{\pi}{\omega} \ln
  \left(\frac{A_H}{4}\right) + S_0 \notag \\
  &=\pi r_+^2+\frac{\pi}{\omega} \ln (\pi r_+^2) -\frac{\pi}{2\omega}
  -\frac{\pi}{\omega}\ln \frac{\pi}{2\omega},
\end{align}
where $A_H=4\pi r_+^2$ is the area of the horizon. The integration
constant $S_0$ was determined in order that the black hole entropy is
non-negative at the minimum mass, which has been eventually chosen as
$S_0 = -\frac{\pi}{2\omega} -\frac{\pi}{\omega}\ln
\frac{\pi}{2\omega}$.  The negative entropy (or zero) has been
discussed in the higher-derivative models~\cite{Cvetic:2001bk}.

The first law of thermodynamics $dE = TdS$ gives the thermodynamic
internal energy
\begin{align}
  E = r + \frac{2}{3\omega r}-\frac{2}{3 \omega r} \left[1 + \omega
    r^2 \left(1 + \frac12 \sqrt{1+ \frac{4M}{\omega r^3}}
    \right)\right] \left[1+ \omega r^2 \left(1- \sqrt{1+
        \frac{4M}{\omega r^3}} \right) \right]^{1/2}, \label{E}
\end{align}
where the integration constant is fixed so that it becomes zero for
$M=0$. Then, it recovers the mass of the black hole at the infinity.
Next, for the critical phenomena, one can calculate the heat capacity
$C$ defined by
\begin{align}
  C &= \frac{\partial E}{\partial T} = T\frac{\partial S}{\partial
    T}, \label{C:def}
\end{align}
where the energy and the entropy are given by Eqs.~\eqref{E}
and~\eqref{bhent}, respectively.  The critical phenomena appear at the
extrema of the local temperature because the energy and the entropy
are monotonic functions with respect to the mass. The logarithmic
correction sometimes gives the negative contribution to the
entropy~\cite{Kaul:2000kf, Govindarajan:2001ee, Chatterjee:2003uv,
  Cai:2009ua}.  Then, the critical phenomena may appear in the extrema
of the energy or the entropy; however, this is not the case in the
present model.

In connection with phase transitions, one can define the free energy,
\begin{align}
  F &= E - TS, \label{F:def}
\end{align}
using the local temperature~\eqref{eq:tem}.  Note that the extrema of
the on-shell free energy are coincident with those of the local
temperature.  On the other hand, the off-shell free energy will be
useful to analyze the stable states of black holes; however, the
meaningful equilibrium occurs at the extrema of the off-shell free
energy.

\section{Critical Phenomena and Phase Transitions}
\label{sec:phenomena}

The thermodynamic behaviors depend on the coupling constant $\omega$,
which means that the KS metric has several black hole states for
$\omega > \omega_{\rm cr}$ and only a single black hole state for
$\omega \le \omega_{\rm cr}$ at a particular temperature.  Note that
the critical value of $\omega_{\rm cr}$ can be found from the
condition that the extrema of the local temperature yield an equal
root.

\begin{figure}[htb]
  \centering
  \includegraphics[width=0.47\textwidth]{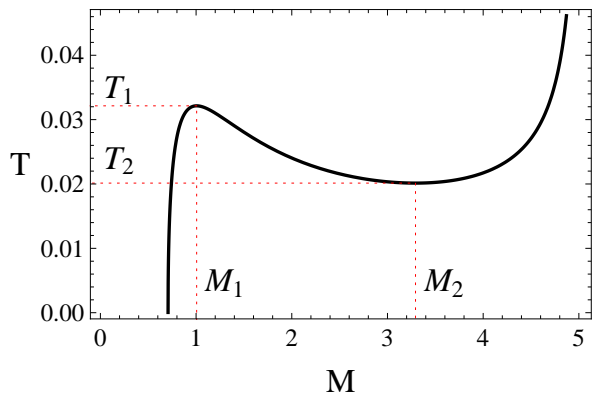}
  \caption{The solid line shows the behavior of the local
    temperature~\eqref{eq:tem} for $\omega > \omega_{\rm cr}$. We set
    $r=10$ and $\omega = 1$. Then, the numerical values become
    $\omega_{\rm cr}=0.215$, $M_1=1.01$, $M_2=3.29$, $T_1=0.0321$, and
    $T_2=0.0201$.}
  \label{fig:largerTem}
\end{figure}

Let us first study for $\omega > \omega_{\rm cr}$.  The black hole
temperature can start with $T=0$ since the black hole may become an
extremal black hole at $M=M_{\rm min}$, and it can have two extrema of
$M_1$ and $M_2$ corresponding to the temperatures $T_1$ and $T_2$ as
seen from Fig.~\ref{fig:largerTem}.  Around $M_1$, let us define the
large black hole for $M >M_1$ and the small black hole for $M < M_1$
for convenience.  We find that the black hole has a single state for
$T > T_1$ or $T < T_2$, three states for $T_2 < T < T_1$, and two
states at $T=T_1$ or $T = T_2$.

\begin{figure}[htb]
  \centering
  \includegraphics[width=0.47\textwidth]{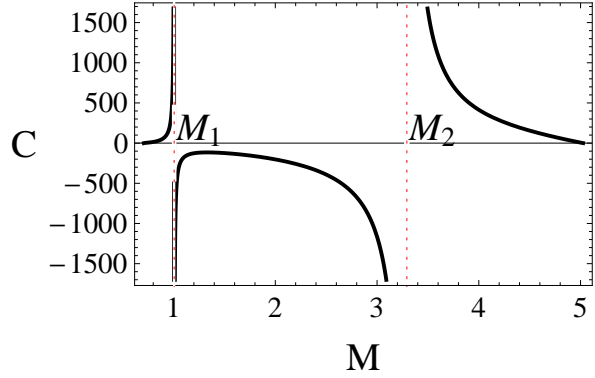}
  \caption{The solid line shows the heat capacity~\eqref{C:def} for
    $\omega > \omega_{\rm cr}$.  This is plotted at $r=10$ and $\omega
    = 1$.}
  \label{fig:largercapa}
\end{figure}

The stabilities of these black holes can be obtained from the heat
capacity~\eqref{C:def}, which is shown in Fig.~\ref{fig:largercapa}.
For $M<M_1$ or $M>M_2$, the black hole is stable whereas it is
unstable for $M_1<M<M_2$.  It means that a large black hole is stable
and the end state after evaporation of the black hole is also stable.
And there exist two stable black holes and an unstable small black
hole simultaneously between $T_2 <T <T_1$.

\begin{figure}[htb]
  \centering 
  \includegraphics[width=0.47\textwidth]{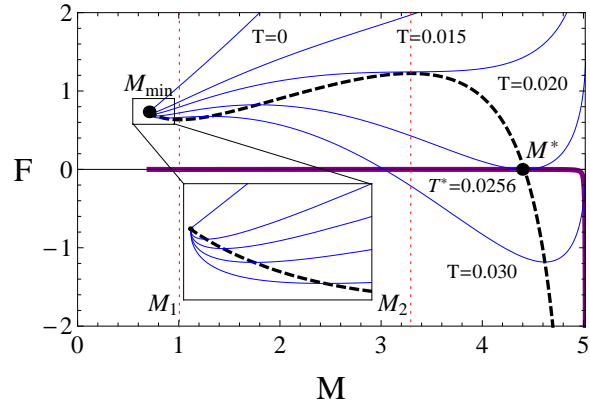} 
  \caption{The thick dashed and solid curves are the on-shell free
    energies of the black hole and the hot flat space,
    respectively. The right black dot is a point of the phase
    transition and a left black dot is a remnant.  The five thin solid curves are the
    off-shell free energies depending on temperature of heat
    reservoir.  These are plotted at $r=10$ and $\omega = 1$ for
    $\omega > \omega_{\rm cr}$. }
  \label{fig:largerfree}
\end{figure}
As for the phase transition from the hot flat space to the stable
large KS black hole, the Hawking-Page type phase
transition~\cite{Hawking:1982dh} appears only above a temperature
$T^*$ as seen from Fig.~\ref{fig:largerfree}.  The free energy of the
hot flat space is $F_{\rm hfs}=-(4/135)\pi^3 r^3 T^4$ which is
actually higher than that of the black hole.  On the other hand, in
the ordinary Schwarzschild black hole, the small unstable black hole
decays either into the large stable black hole or to the thermal
state; however, it can decay into the small stable black hole in this
KS solution. Note that a stable black hole exists at $0<T<T_1$.  It is
locally stable, so that it should undergo a quantum tunneling and
decay to the hot flat space below the critical temperature $T^*$ as
seen from Fig.~\ref{fig:largerTem} and Fig.~\ref{fig:largerfree}. And,
the remnant whose mass is $M_{\text{min}}$ should decay into the
vacuum through the quantum tunneling. At $T = 0$, it disappears and
becomes an extremal black hole, which is no longer thermodynamic
object.  The reason why the on-shell and the off-shell free energies
at $M=M_{\rm min}$ are all the same $F_{\rm on} =F_{\rm off} =E$ is
that the entropy is zero for any temperature. This extremal
configuration is similar to the case of the noncommutative black holes
from the profile of the temperature~\cite{Kim:2008vi,
  Nicolini:2011dp}.

\begin{figure}[htb]
  \centering
  \includegraphics[width=0.47\textwidth]{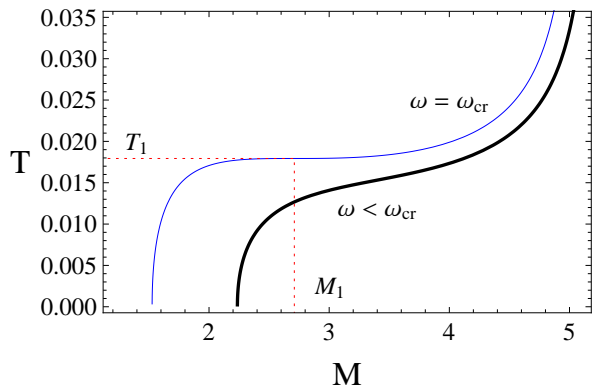}
  \caption{The thin and the thick solid lines show the local
    temperatures for $\omega=\omega_{\rm cr}$ and $\omega <
    \omega_{\rm cr}$, respectively. There does not exist any extremum
    at all. This is plotted for $r=10$ and $\omega=0.1$. Then, the
    numerical values are given by $\omega_{\rm cr}=0.215$, $M_1=2.71$,
    and $T_1=0.0179$.}
  \label{fig:smallerTem}
\end{figure}

\begin{figure}[h]
  \centering
  \includegraphics[width=0.45\textwidth]{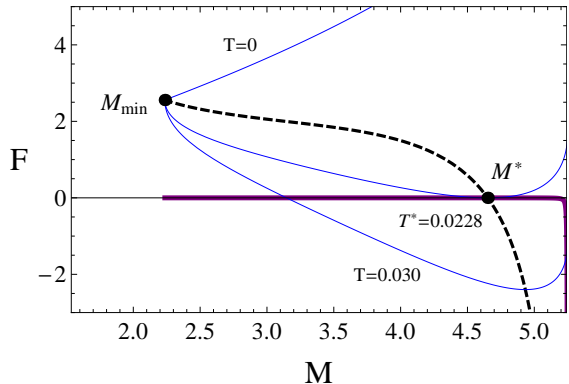} 
  \caption{The thick dashed and solid curves are the on-shell free
    energies for the black hole and for the hot flat space,
    respectively for $\omega < \omega_{\rm cr}$. The phase transition
    occurs at $T^*$. The plot for $\omega =\omega_{\rm cr}$ is very
    similar to this case. This is plotted for $r=10$ and $\omega =
    0.1$, which results in $\omega_{\rm cr} =0.215$, $M^*=4.65$, and
    $T^*=0.0218$.}
  \label{fig:smallerfree}
\end{figure}

Let us now consider the case of $\omega \le \omega_{\rm cr}$.  The
local temperature \eqref{eq:tem} monotonically increases with respect
to the mass as shown in Fig.~\ref{fig:smallerTem}.
Any black hole state is stable since the heat capacity is always
positive. Specifically, the maximum of the heat capacity appears at
$M_1$ for $\omega < \omega_{\rm cr}$, and it is positively divergent
at that point for $\omega =\omega_{\rm cr}$. Especially, the heat
capacities vanish at $M= M_{\rm min}$ and $M =M_{\rm max}$.
In this case also, there appears the Hawking-Page type phase
transition since the on-shell free energy of the hot flat space is
higher than that of the black hole, so that the stable black hole is
more preferable above the critical temperature $T^*$ as seen from
Fig.~\ref{fig:smallerfree}.  For usual phase transitions, it seems to
be necessary that the black hole mass should be getting smaller and
smaller and then eventually its mass disappears to decay completely to
the hot flat space without the black hole. On the contrary to the
ordinary Schwarzschild black hole which has a continuous mass
spectrum, the KS black hole has a lower bound. At first sight, the
black hole does not decay into the hot flat space below $T^*$;
however, this is not the case since the on-shell free energy of the
black hole is higher than the free energy of the hot flat space below
the critical mass $M^*$ as seen from Fig.~\ref{fig:smallerfree}, so
that it should decay into the hot flat space quantum-mechanically.

As a result, the Hawking-Page phase transition from the black hole to
the hot flat space can occur at a critical temperature for any
coupling constant of $\omega$. For $\omega > \omega_c$, even though
there exists a locally stable small black hole below the critical
temperature, it should tunnel to the large black hole which eventually
decays into the hot flat space. For a very high temperature, it
directly decays into the large black hole thermodynamically.  For
$\omega \le \omega_c$, all black hole states are stable and they
tunnel to the hot flat space quantum-mechanically below the critical
temperature.  Finally, the remnant should also tunnel to the vacuum.

\section{Discussion}
\label{sec:discus}

We have studied the thermodynamics and the phase transitions of the KS
black hole at the finite isothermal surface.  Note that in this model
the mass of the black hole should be bounded.  Especially, the lower
bound of the mass is related to the remnant and the upper bound comes
from the cavity effect.  The KS black hole metric depends on a
coupling constant $\omega$ which gives inner and outer horizons. When
two horizons are coincident, it gives the minimum mass.  By the way,
the black hole is surrounded by the cavity, that is, the radius of the
outer horizon should be smaller than the size of cavity. When the size
of cavity equals to the outer horizon, the black hole has a maximum
mass. So, the temperature of the black hole with the minimum mass is
zero from the extremality while the temperature at the maximum mass is
infinity. Between them, the extrema have something to do with the
critical phenomena.  For the range of $\omega \le \omega_{\rm cr}$,
the temperature has no extremum and monotonically increases as seen
from Fig.~\ref{fig:smallerTem}. For $\omega > \omega_{\rm cr}$, there
are two extrema which are related to the stability of the black hole
states.

On general ground, a lower on-shell free energy state is more
preferable so that Hawking-Page type phase transition from the hot
flat space to the black hole naturally appears for the large black
hole. As for the small black hole, a locally unstable small black hole
can decay into the large black hole or into the locally stable small
black hole thermodynamically below the critical temperature.
Subsequently, the stable small black hole tunnels to the hot flat
space quantum-mechanically as seen from Fig.~\ref{fig:largerfree} and
Fig.~\ref{fig:smallerfree}.

Finally, let us discuss the evolution of a small unstable black hole
especially for $\omega > \omega_{\rm cr}$. Suppose that it loses its
mass through Hawking radiation, then the negative heat capacity shows
that the black hole temperature should be increased while the
temperature of the heat reservoir is still maintained. Eventually, it
can decay into much smaller stable black hole which lies in the lower
free energy state compared to that of the unstable black hole.  In
other words, it arrives at the on-shell state of the same temperature
with that of the heat reservoir since there are two black hole states
for a given temperature.  If this black hole loses its mass further,
it subsequently acquires some energy from the heat reservoir because
of the positive heat capacity. Even though it can arrive at the
minimum mass throughout the non-trivial mechanism, it can not evolve
any more.  Since the extremal black hole with the non-vanishing
temperature of the heat reservoir is not in thermal equilibrium, so
that it should return to the stable black hole eventually.  Then, the
stable small black hole should decay into the thermal vacuum via a
quantum tunneling.

\begin{acknowledgments}
  We would like to thank S.-H. Yi for exciting discussion.  M.~Eune
  was supported by National Research Foundation of Korea Grant funded
  by the Korean Government (Ministry of Education, Science and
  Technology) (NRF-2010-359-C00007). B.~Gwak and W.~Kim are supported
  by the National Research Foundation of Korea(NRF) grant funded by
  the Korea government(MEST) through the Center for Quantum
  Spacetime(CQUeST) of Sogang University with grant number
  2005-0049409, and W.~Kim was also supported by the Basic Science
  Research Program through the National Research Foundation of
  Korea(NRF) funded by the Ministry of Education, Science and
  Technology(2012-0002880).
\end{acknowledgments}



\end{document}